# Structure and Magnetism of the mono-layer hydrate $Na_{0.3}NiO_2$ $0.7H_2O$

S. Park[a], W.-S. Yoon[b], T. Vogt[a,]*

[a]USC NanoCenter & Department of Chemistry and Biochemistry, University of South Carolina, 1212 Greene Street, Columbia, SC 29208-0001, USA
[b]Chemistry Department, Brookhaven National Laboratory, Upton, NY 11973-5000, USA

The mono-layer hydrate (Ni-MLH) $Na_{0.3}NiO_2$ $0.7H_2O$ was synthesized with an average Ni valence close to one in the bi-layer hydrate (Ni-BLH) $Na_{0.3}NiO_2$ $1.3H_2O$. A weak unsaturated ferromagnetism and divergence of the magnetic susceptibilities of field and zero-field cooled samples of both Ni-MLH and Ni-BLH are observed. However, as a result of the increased 3-dimensional electronic and magnetic character present in Ni-MLH a frequency dependence of the AC magnetic susceptibility indicative of the spin glass-like behavior as seen in Ni-BLH was no longer observed in Ni-MLH.

# 1. Introduction

The discovery of low temperature superconductivity (Tc<5K) in $Na_{0.3}CoO_2$ $1.3H_2O$ [1] has refocused scientific attention to the $A_xMO_2$ $nH_2O$ family of 2-dimensional oxide hydrates. The bilayer hydrate (BLH) structure of this superconducting phase has a water-Na-water layer separating the doped $CoO_2$ layers. In the monolayer hydrate (Co-MLH) $Na_{0.36}CoO_2$ $0.7H_2O$ the 'charge reservoir' bilayer is replaced by a single layer containing both Na and water. Superconductivity above 2K is no longer found in Bi-MLH. The suppression of superconductivity in Co-MLH has been the subject of various studies. Early work by Sakurai [2] pointed to an effective shielding of the random Coulomb potential of the Na+ cations by the external water layers in the BLH structure resulting in a more homogeneous potential present in the $CoO_2$ layers. First principles electronic band structure calculations by Arita [3] revealed that the dispersion of the $a_{1g}$ band near the Fermi level along the c-axis is negligible in the Co-BLH phase but on the order of 1/10 eV in the Co-MLH phase. It appears thus that augmenting the 3-dimensional character of the $a_{1g}$ band might play a role in suppressing superconductivity. Recent angle-resolved photoemission studies underline the importance of the a1g band [4].

While other layered metal oxide systems in the $A_xMO_2$ $nH_2O$ family have also been shown to crystallize in either bilayer (three layer BLH $Na0.3CoO_2$ $1.3H_2O$ [5], $Na0.3NiO_2$ $1.3H_2O$ [6] or monolayer hydrates ($Na_{0.36}CoO_2·0.7H_2O$[7], $K_{0.3}CoO_2·0.4H_2O$[8], $Na_{0.3}RhO_2·0.6H_2O$ [9], $Na_{0.22}RuO_2·0.45H_2O$ [10], and $Na_{0.32}MnO_2·0.67H_2O$ [11] ) only for A=Na, M=Ni and A=Na, M=Co do both a BLH and a MLH exist (for x~0.3 and n=1.3 and 0.7). The crystal structure of $Na_{0.3}CoO_2·1.3H_2O$ (*P6₃/mmc*) is built up of hexagonal layers of 30% nonmagnetic $Co^{3+}$ and 70% low-spin $Co^{4+}$ (S=1/2) ions. The crystal structure of $Na_{0.3}NiO_2·1.3H_2O$ has a monoclinic symmetry (*C2/m*) where 30% low-spin $Ni^{3+}$ (S=1/2) and 70% nonmagnetic

$Ni^{4+}$ ions decorate the triangular lattice. Formally the oxidation and hydration of $NaCoO_2$ to $Na_{0.3}CoO_2 \cdot 1.3H_2O$ results in the formation of a majority of S=1/2 ions, whereas in the case of the $Na_{0.3}NiO_2 \cdot 1.3H_2O$ system the s=1/2 minority ions are diluted by nonmagnetic ions. The synthesis and structure of the Ni-BLH (x=0.3, n=1.3) was recently described 6. The magnetic properties of Ni-BLH revealed a spin glass-like behavior at temperatures below 5K [12]. The shorter $NiO_2$ inter-layer separation and Ni-O-Na-O-Ni inter-layer superexchange pathssimilar to those in $NaNiO_2$ possible in Ni-MLH compared to Ni-BLH should result in changes of the magnetism. It appears thus interesting to characterize and compare the physical properties of Ni-BLH and Ni-MLH.

## 2. Experimental

Powder samples of mono-layer hydrate $Na_{0.3}NiO_2 \cdot 0.7H_2O$ (MLH) and bi-layer hydrate $Na_{0.3}NiO_2 \cdot 1.3H_2O$ (BLH) were prepared from $Na_xNiO_2$ by extracting $Na^+$ cations and simultaneously intercalating water using $Na_2S_2O_8$ as shown in figure 1. Detailed synthesis procedure was described previously [13]. High-resolution synchrotron X-ray powder diffraction data was measured at beam line X7A at the National Synchrotron Light Source at Brookhaven National Laboratory employing a gas-proportional position-sensitive detector (PSD), using a gated Kr-escape peak [14]. The Ni K-edge X-ray absorption spectroscopy (XAS) experiments were performed on beam line X11A at the National Synchrotron Light Source. A pair of Si (111) crystals was used as a monochromator and harmonic contaminations were suppressed by appropriate detuning. A Ni metal foil was measured simultaneously in every experiment to calibrate the energy scale. Magnetization measurements were carried out using powdered samples in a superconducting quantum interference device (SQUID) magnetometer with both DC and AC fields. For zero-field studies the superconducting magnet was demagnetized prior to data collection. Temperature steps of 0.02 K were used. The magnetizations M versus the applied field H curves were measured at 30 and 2 K. The DC magnetic susceptibility *M/H* was measured following two procedures. The first procedure involved cooling the sample from 30 K down to 2 K in nominal zero field and then measuring the magnetization under various fields in the range 2 Oe ≤ H ≤ 1000 Oe while heating up to 30 K. This is referred to as zero-field-cooling (ZFC) magnetization. The second procedure involved measuring the magnetic moment while cooling from 30 K down to 2 K under a constant applied field, which is referred to as field-cooling (FC) magnetization. The AC magnetic susceptibility was measured in the frequency range 1 - 100 Hz with an AC driving field of 1 Oe at a dc field level of 0 Oe.

## 3. Results and discussion

The oxidation and concomitant water intercalation into $NaNiO_2$ using $Na_2S_2O_8$ in an aqueous solution leads to the appearance of first $Na_{0.3}NiO_2 \cdot 0.7H_2O$ (Ni-MLH) and then $Na_{0.3}NiO_2 \cdot 1.3H_2O$ (Ni-BLH). The $Na_{0.3}NiO_2 \cdot 0.7H_2O$ (Ni-MLH) structure where $H_2O$ molecules and sodium cation sites are contained within a single layer between $NiO_2$ layers is shown in Figure 1. The $NiO_2$ layers are made of edge-sharing $NiO_6$ octahedra forming a triangular $MO_2$ lattice. The structure of $Na_{0.3}NiO_2 \cdot 1.3H_2O$ (Ni-BLH) consists of layers of nickel oxide octahedra separated by layers of $H_2O$-Na-$H_2O$ molecules. In both $NaNiO_2$ and Ni-MLH similar Ni-O-Na-O-Ni magnetic exchange paths exist along the c-axis. Due to the presence of water in the 'charge reservoir' layer of Ni-MLH these interactions are expected to be weaker than in $NaNiO_2$. In the Ni-BLH structure however, this exchange path is even more weakened by the disordered 3 layer 'charge reservoir'. The electronic and magnetic structure is thus expected to be more 2-dimensional.

Synchrotron powder X-ray diffractions of both Ni-BLH and Ni-MLH revealed powder patterns consistent with a monoclinic cell with axis lengths of a = 4.9023(3)Å b = 2.8267(1) Å for Ni-MLH and a = 4.9064(1) Å b = 2.9214(3) Å for Ni-BLH, which are different from those a = 5.3226(9) Å, b = 2.8451(5) Å of $NaNiO_2$ as shown in Table 1 due to the different Ni valency and water intercalation.. The monoclinic angles (β = 103.573(6)° for Ni-MLH and β = 108.074(3) ° for Ni-BLH are slightly smaller than the one observed in the parent compound $NaNiO_2$ (β = 110.493(3) °). The $NiO_2$ interlayer spacing of the parent compound $NaNiO_2$ is 5.2 Å. This interlayer spacing expands to 7.1 Å in the Ni-MLH and 10.0 Å in the Ni-BLH as shown in Figure 2 by the shift of the 00l reflections to lower Bragg angles and are similar to the $CoO_2$ interlayer distances of 6.9 Å and 9.8 Å observed in Co-MLH [15] and Co-BLH [16] respectively.

Figure 3 shows the normalized Ni K-edge XANES spectra for a Ni metal foil, $Na_xNiO_2$ (precursor), $Na_{0.3}NiO_2 \cdot 0.7H_2O$, and $Na_{0.3}NiO_2 \cdot 1.3H_2O$. The Ni K-edge XANES spectra originate from transitions of the 1s core electrons of the Ni ions to excited vacant bound states of proper symmetry. It can be seen readily that as Na is deintercalated and the structure is hydrated that the entire edge position shifts to higher energies in both Ni-MLH and Ni-BLH. These shifts in the edge-position indicate that at least formally the average valence of Ni increases. However, as suggested in previous work [6] this can be due to Ni being in a combination of $3d^7$ (s=1/2) and $3d^6$ (s=0) or $3d^7L$ states, with L representing a hole in the oxygen p-states. This approach has shown to reproduce 2p-x-ray absorption spectra of $NaNiO_2$ better than the simple atomic-multiplet calculation [17]. Classically one assumes a low spin S=1/2 $Ni^{3+}$ ion to be present in $NaNiO_2$. However, a spin composite of a S=1 $Ni^{2+}$ state bound to an s=1/2 hole resulting in an overall s=1/2 Zhang-Rice singlet as advocated above might be a better description. The Ni absorption spectra do show that the average Ni valence in both Ni-MLH and Ni-BLH is very close. This is similar to what is observed in Co-BLH and Co-MLH where x-ray absorption spectroscopy studies reveal that there is essentially no change of the Co valence upon hydration and that any role of the water acting as a supplementary charge reservoir source is minor [18].

Figure 4 (a) shows the partial demagnetization curves for Ni-MLH measured at 30 K and at 2 K. While the 30 K curve shows paramagnetic behavior, the 2 K curve exhibits an unsaturated ferromagnetism with a small remnant magnetization (inset, Fig. 4 (a)). A similar paramagnetic and unsaturated ferromagnetism behavior at 30 K and at 2 K is observed in Ni-BLH [5]. Figure 4 (b) shows a typical temperature dependence of the magnetic susceptibility of Ni-MLH with an applied field of $H_{appl}$ = 2 and 100 Oe. A divergence of the FC and the ZFC measurements is observed at 2 Oe with the FC magnetization increasing monotonically with decreasing temperature and the ZFC one exhibiting a maximum magnetization near 2.5 K. In contrast both the FC and ZFC susceptibility increase with decreasing temperature in a field of 100 Oe. The inset in Figure 4 (b) shows the magnetic susceptibility as a function of temperature for Ni-BLH.

The divergence of the FC and the ZFC measurements is observed at higher temperatures and an applied field of $H_{app}$ =100 Oe.

AC susceptibility measurements performed as a function of temperature $T$ and frequency ν at a constant dc field level $H_{dc}$ were performed to probe for a spin-glass-like state. The frequency-dependent AC susceptibility allows the determination of the freezing temperatures ($T_{sf}$), which is the mean relaxation rate. The existence of such a small remnant magnetization below a freezing temperature $T_{sf}$ is generally observed in spin-glasses [19]. However, such an observation is a necessary but not sufficient condition for a 'true' spin glass. After ZFC to 2 K, $\chi'(T)$ and $\chi'(\nu)$ were recorded (Figures 4 (c)). In general the AC susceptibility curves show a relatively broad maximum. The inset of Figure 4 (c) shows the decrease of $\chi(ac)_{max}$ with increasing frequency at zero dc applied field for Ni-BLH. In contrast to $\chi(ac)_{max}$ of Ni-BLH depending on temperature and frequency at zero dc applied field $\chi(ac)_{max}$ for Ni-MLH reveals no longer a frequency-dependent component as shown in Figure 4c. The loss of the frequency dependence of $\chi(ac)_{max}$ is an indication that Ni-MLH has lost its spin glass-like characteristics.

In the parent compound $NaNiO_2$, which shows orbital ordering at 480K and in-plane long-range ferromagnetism below 20K [20], the antiferromagnetic exchange between $NiO_2$ layers (~-0.1 meV) is considerable weaker than the ferromagnetic exchange within the $NiO_2$ layers (~1 meV) [21]. As mentioned earlier the antiferromagnetic interactions are mediated via quite long Ni-O-Na-O-Ni superexchange paths running at roughly 45 degrees in the ac-plane, while the ferromagnetic interactions are caused by 90º Ni-O-Ni exchange following the Goodenough-Kanamori-Anderson rules [22]. The actual Ni-O-Ni bond angles are ~95º at room temperature. Recent muon-spin rotation, heat capacity and magnetic measurements of $NaNiO_2$ reveal a significant presence of short-range magnetic order between 5K and $T_N$=19.5K[23]. Similar to what was observed in Ni-BLH a frequency dependent peak in the ac magnetic susceptibility near 3K was observed in $NaNiO_2$, indicative of a significant slowing down of spin fluctuations. Fitting the frequencies and spin transition temperatures in $NaNiO_2$ to a power law T= $T_{sf}$ [1+(fτ$_0$}]$^{1/zv}$ resulted in $T_{sf}$ = 3.0(2)K , τ$_0$=5.4(2) 10$^{-3}$ s and zv= 8.1(4). The frequency dependence of the ac magnetic susceptibility measured in Ni-BLH and analyzed in the same way resulted in very similar parameters: $T_{sf}$=4.4K, $\tau_0$=4.7 $10^{-3}$s and zv=4.03 [6]. Spin glass-like properties have also been observed in $Li_{1-x}Ni_{1+x}O_2$ samples with x~0.01 [24]. However, due to the difficulties in synthesizing stoichiometric $LiNiO_2$ sample-dependent results plague the research literature. In $NaNiO_2$ the frequency dependence of the ac magnetic susceptibility excluded a 'true' spin glass since muon precession indicative of magnetic long range order was observed down to 1.6K. Instead the authors argue that either a small concentration of oxygen vacancies associated with two $Ni^{2+}$ impurities form ferromagnetic clusters within the $NiO_2$ layers or simply a significant slowing down of spin fluctuations occurs at $T_{sf}$. Our observation of a similar magnetic behavior in Ni-BLH points to a mainly intra-layer origin of this spin glass-like behavior since the exchange interactions along the c axis in the BLH structure would be extremly weak. The fact that this frequency-dependent behavior is not present in Ni-MLH is intriguing and not yet understood. The frequency-independent peak observed in Ni-MLH at ~2.5K could be of the same origin as the one observed just slightly above 3K in Ni-BLH (see inset Figure 4c). We had tentatively assigned this peak to an impurity present in the Ni-BLH sample. While it is of course possible that the impurity is present in both samples this peak could also have intrinsic origins in both Ni-BLH and Ni-MLH.

**Acknowledgements**

The authors thank the National Synchrotron Light Source at Brookhaven National Laboratory.

**Figure captions :**

**Figure 1.** Structures of $NaNiO_2$ (ICSD 415072), $Na_{0.3}NiO_2 \cdot 0.7H_2O$ (MLH), and $Na_{0.3}NiO_2 \cdot 1.3H_2O$ (BLH). Lines in the figures define the unit cells.

**Figure 2.** Calculated X-ray diffraction pattern for (a) $NaNiO_2$ (ICSD 415072) and observed X-ray diffraction patterns for (b) $Na_{0.3}NiO_2 \cdot 0.7H_2O$ (MLH) and (c) $Na_{0.3}NiO_2 \cdot 1.3H_2O$ (BLH).

**Figure 3.** The normalized Ni K-edge XANES spectra of Ni metal foil, precursor $Na_xNiO_2$, $Na_{0.3}NiO_2 \cdot 0.7H_2O$ (Ni-MLH), and $Na_{0.3}NiO_2 \cdot 1.3H_2O$ (Ni-BLH).

**Figure 4.** (a) Partial demagnetization curves of $Na_{0.3}NiO_2 \cdot 0.7H_2O$ (Ni-MLH) measured at 30 and at 2 K. The inset shows the magnified curves in a low magnetic field. (b) Typical temperature dependence of the susceptibility of $Na_{0.3}NiO_2 \cdot 0.7H_2O$ (Ni-MLH) with $H_{app}$ = 2 and 100 Oe. The inset shows a typical temperature dependence of the susceptibility of $Na_{0.3}NiO_2 \cdot 1.3H_2O$ (Ni-BLH) with $H_{app}$ = 100 Oe. (c) AC susceptibilities of $Na_{0.3}NiO_2 \cdot 0.7H_2O$ (Ni-MLH) as a function of temperature $T$ and frequency ν at constant DC field. The inset shows AC susceptibilities of $Na_{0.3}NiO_2 \cdot 1.3H_2O$ (Ni-BLH) as a function of temperature $T$ and frequency ν at constant DC field.

**Figure 1.**

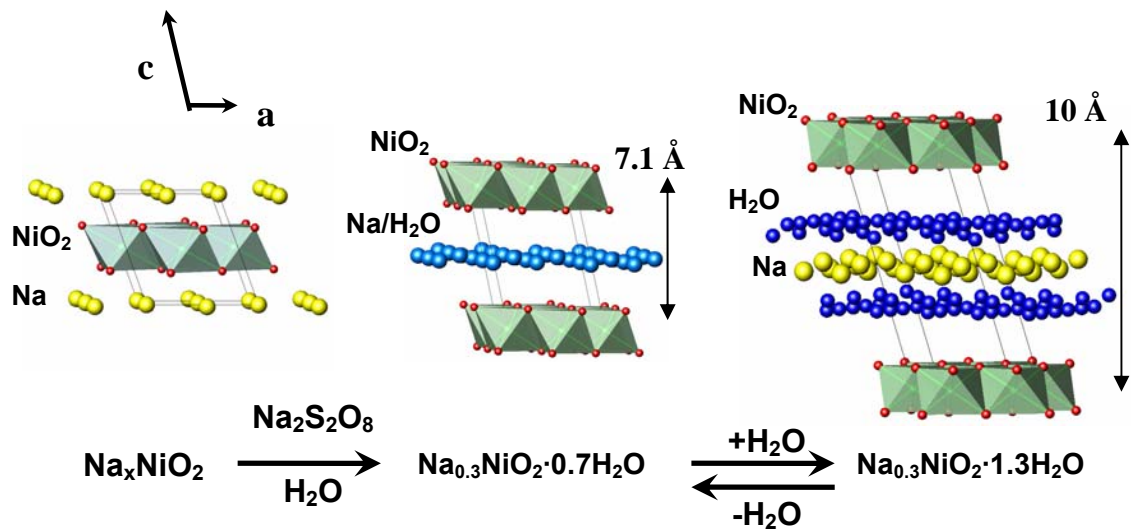

**Figure 2.**

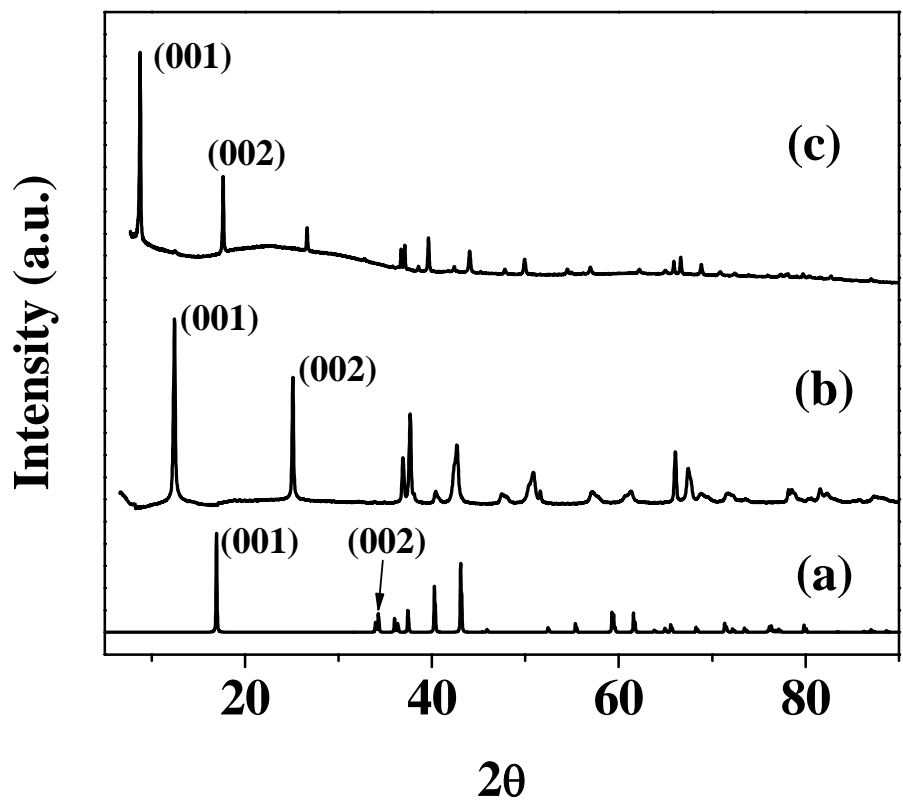

**Figure 3.**

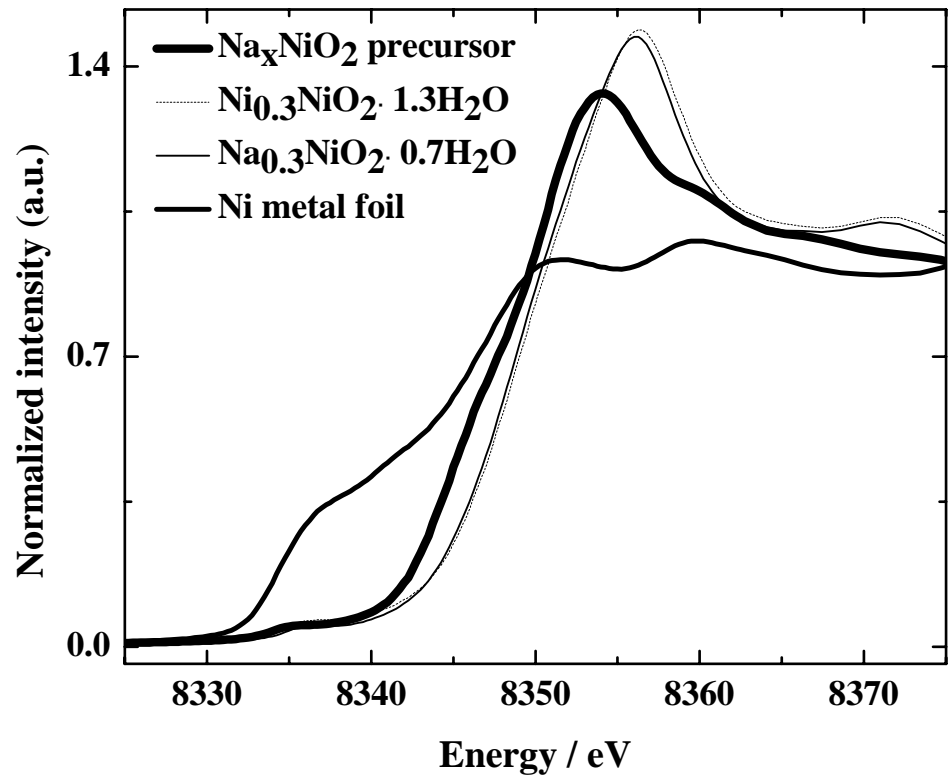

**Figure 4a**

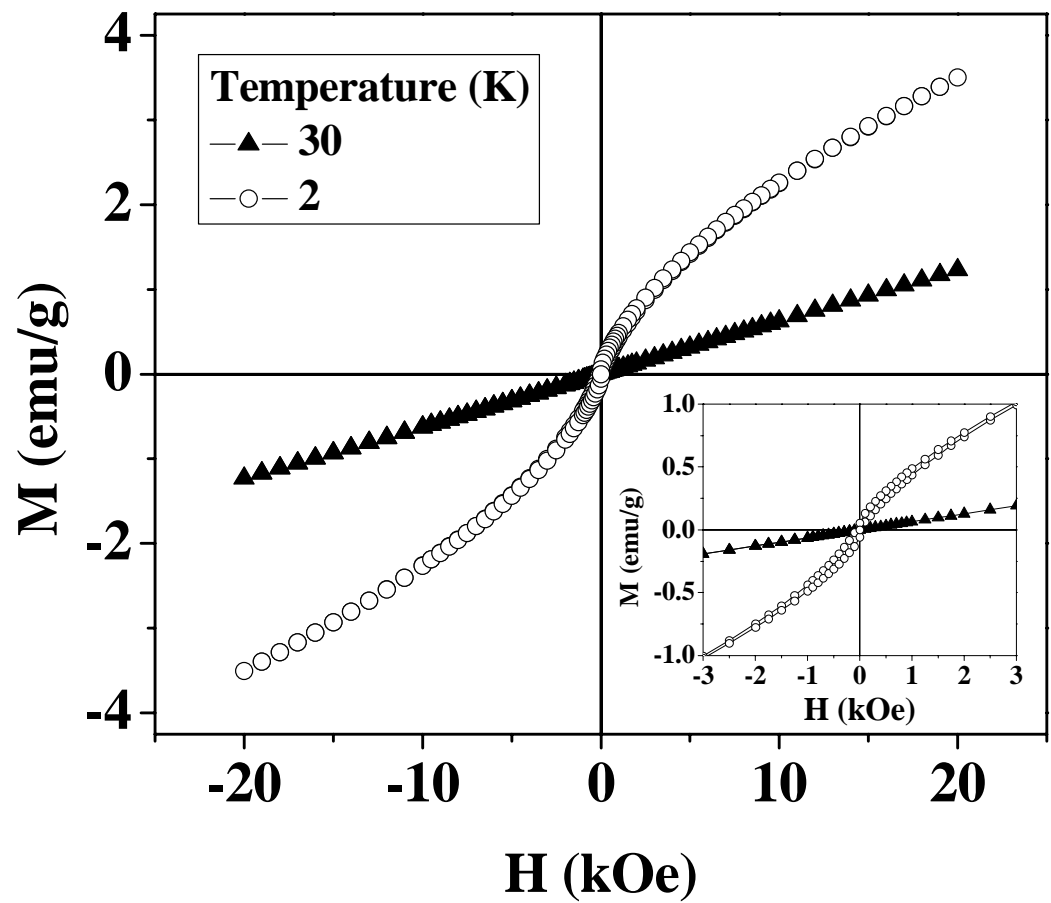

**Figure 4b**

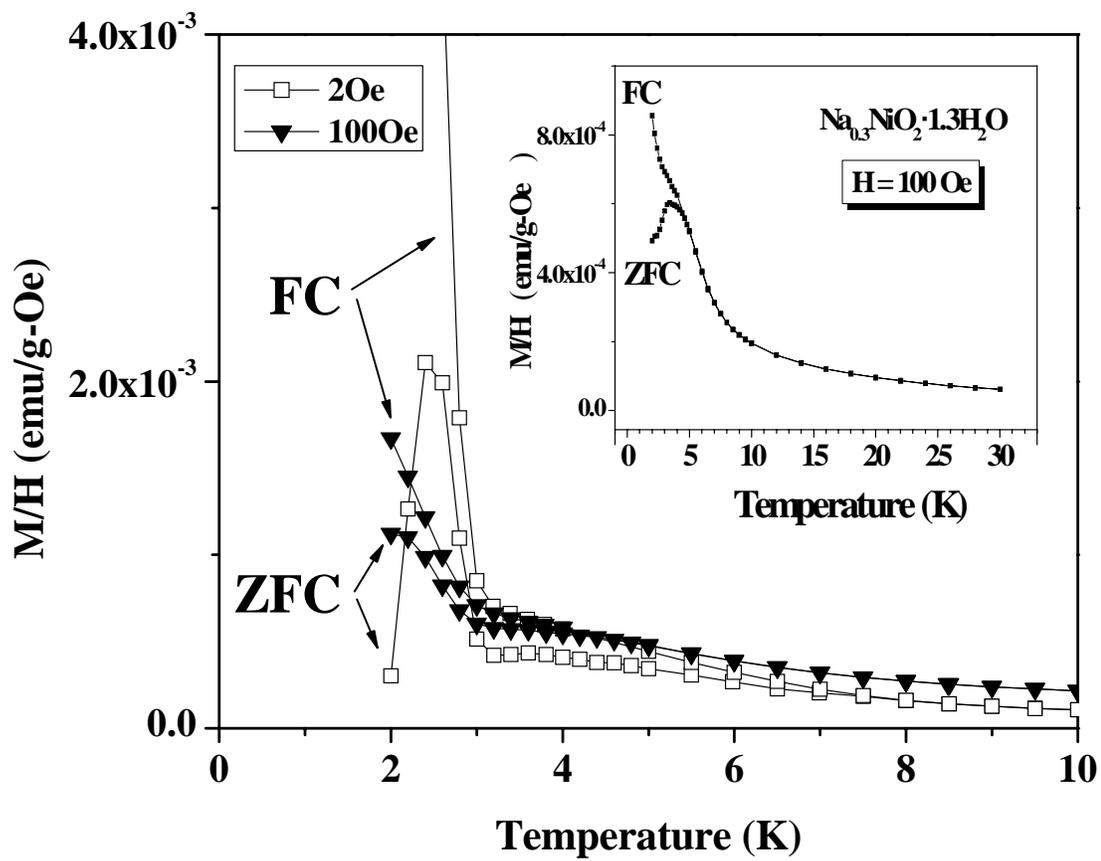

**Figure 4c**

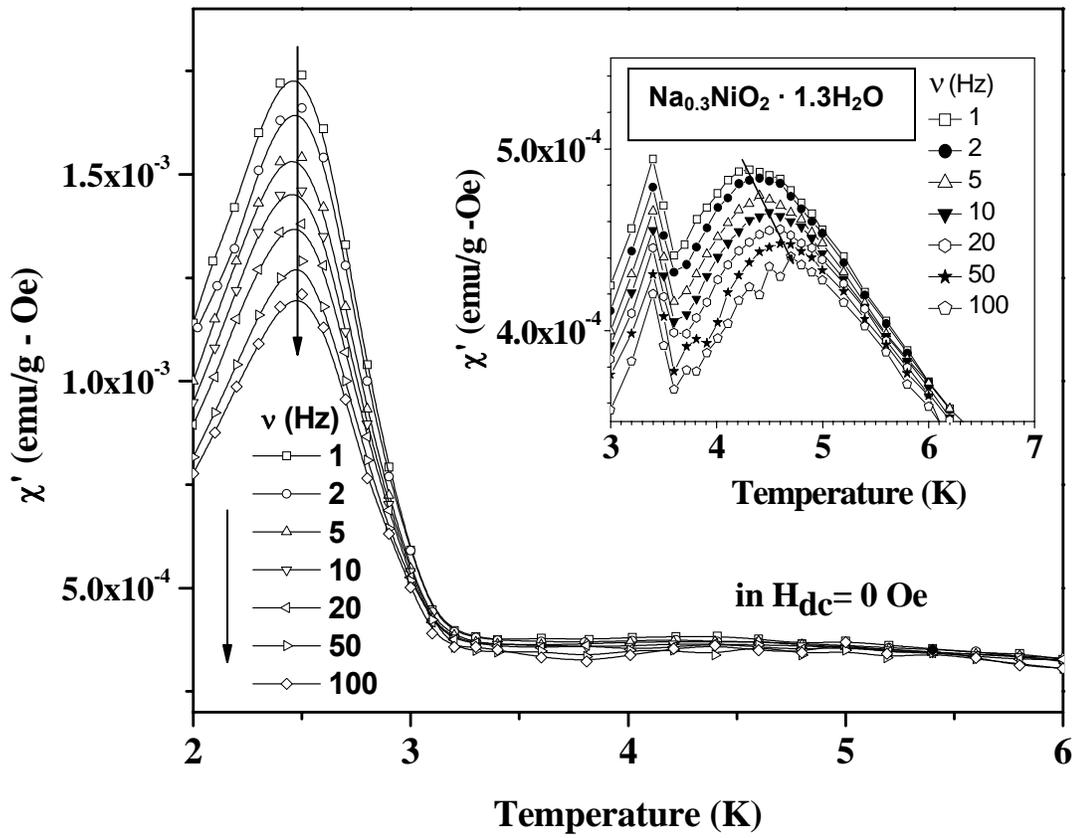

**Table 1. Unit cell, volume, and interlayer spacing calculations for $NaNiO_2$, $Na_{0.3}NiO_2 \cdot 0.7H_2O$, $Na_{0.3}NiO_2 \cdot 1.3D_2O$, and $Na_{0.3}NiO_2 \cdot 1.3H_2O$.**

|  | Unit Cell (Å) | | | | Volume ($Å^3$) | Interlayer Spacing(Å) |
|---|---|---|---|---|---|---|
|  | a | b | c | β | | |
| $Na_{0.3}NiO_2 \cdot 1.3H_2O$ | 4.9064(1) | 2.9214(3) | 10.5348(6) | 108.074(3) | 143.6 | 10.0 |
| [1]$Na_{0.3}NiO_2 \cdot 1.3D_2O$ | 4.890(2) | 2.9361(9) | 10.542(2) | 107.81(2) | 144.1 | 10.0 |
| $Na_{0.3}NiO_2 \cdot 0.7H_2O$ | 4.9023(3) | 2.8267(1) | 7.2786(5) | 103.573(6) | 98.05 | 7.1 |
| [2]$NaNiO_2$ | 5.3226(9) | 2.8451(5) | 5.5842(10) | 110.493(3) | 79.21 | 5.2 |

[1] Ref. [6]   [2] ICSD 415072